\documentclass[preprints,article,accept,moreauthors,pdftex,10pt,a4paper]{mdpi}

\firstpage{1} 
\pubvolume{xx}
\issuenum{1}
\articlenumber{5}
\pubyear{2018}
\copyrightyear{2018}
\history{Submitted: 2018. 10. 31.}

\continuouspages{yes}
\pdfoutput=1
\usepackage{amssymb}
\Title{Non-extensive Motivated Parton Fragmentation Functions$^{\dagger}$}

\Author{Adam Takacs $^{1,2}$\orcidA{} and Gergely G\'{a}bor Barnaf\"{o}ldi $^{2}$\orcidB{}}

\AuthorNames{Adam Takacs, Gergely G\'{a}bor Barnaf\"{o}ldi}

\address{%
$^{1}$ \quad Institute of Physics, E\"{o}tv\"{o}s University, 1/A P\'azm\'any P\'eter s, H-1117 Budapest, Hungary;\\
$^{2}$ \quad Wigner Research Centre for Physics of the H.A.S., P.O. Box, H-1525 Budapest, Hungary;}

\corres{Correspondence: takacs.adam@wigner.mta.hu; barnafoldi.gergely@wigner.mta.hu;}

\firstnote{Presented at Hot Quarks 2018 - Workshop for young scientists on the physics of ultrarelativistic nucleus-nucleus collisions, Texel, The Netherlands, September 7-14 2018}

\abstract{A new form of fragmentation function is presented here, motivated by earlier non-extensive studies of jet fragmentation. We parametrized our Tsallis-like function on pion spectra and compared it to the most common fragmentation function parametrizations. It is shown that the new form is in agreement with earlier parametrizations, furthermore, its scale evolution overlap better with the experimental data.}
\keyword{parton model; fragmentation; hadronization; QCD; hadron spectrum; non-extensive statistics; DGLAP evolution; Tsallis distribution; $e^-e^+$ collision; $pp$ collision; heavy ion collision;}

\begin{document}

\section{Introduction}

In high energy collisions the energy distribution of the outgoing particles with low momentum ($\lesssim3$ GeV/c) follows a thermal-like, exponential distribution. Particles with higher momentum ($\gtrsim5$ GeV/c), where pQCD works better, follow power law distributions. The whole spectrum cannot be described by one exponential or power law function, but the Tsallis distribution fits it perfectly~\citep{Biro:2017arf}.

Statistical systems, like high-energy proton-proton ($pp$) or electron-positron ($e^-e^+$) collisions are small. The created $\sim$\,100 particles are far from the thermodynamic limit, highly fluctuating and correlated. Therefore, they could be beyond the validity of statistical mechanics, which was built for weakly interacting systems~\cite{tsallis2009introduction,campa2014physics,Takacs:2017wnn}. The $q$ entropy is a generalization of Boltzmann's one, which could lead to a better description of this fluctuating collision system~\citep{Biro:2014fluct}. The theory has a temperature-like parameter, $T$ and an additional $q$ parameter describing the non-additivity of the entropy for independent systems. Tsallis is the characteristic distribution, derived from this so-called non-extensive entropy, analogue to the Boltzmann distribution. In the $q\rightarrow1$ limit, Tsallis' entropy and distribution restores Boltzmann's.

In Ref.\cite{Urmossy:2011xk}, we built a model for jet fragmentation resulting the so-called microcanonical Tsallis distribution, derived from the observed Gamma shape of the multiplicity distribution. This result is extended here, by applying the form as a fragmentation function (FF). Our paper has two aims: (i) to present the validity of our FF assumption by comparing it to widely applied FFs and experimental data. (ii) To investigate the non-extensive parameters of FFs by comparing them to earlier studies~\citep{Biro:2017arf}.

\section{The Formalism}

In the parton model assuming the factorization, perturbative (e.g. partonic cross sections) and non-perturbative (e.g. parton ditribution and hadronization) processes are treated separately. The normalized cross section of the outgoing $h$ hadrons in $e^-e^+$ annihilation is defined by in~\citep{ellis2003qcd,Patrignani:2016xqp},
\begin{equation}\label{eq:total_fragmentation}
F^h(z,Q^2)=\frac{1}{\sigma_{tot}}\frac{\mathrm{d}\sigma(e^-e^+\rightarrow hX)}{\mathrm{d}z},
\end{equation}
where $Q$ is the momentum scale, $z=2E_h/\sqrt{s}$ is the momentum fraction, where hadron energy is $E_h$ and $\sqrt s$ is the center of mass energy. The $\sigma_{tot}(Q)$ is the total cross section. The inclusive differential cross section of hadron production is
\begin{equation}\label{eq:differential_fragmentation}
\frac{\mathrm{d}\sigma^h}{\mathrm{d}z}=\sum_i\int^1_z\frac{\mathrm{d}x}{x}C_i(x,\alpha_s,Q^2)D^h_i(z/x,Q^2),
\end{equation}
where $C_i$ functions are calculated from perturbative QCD. The non-perturbative $D^h_i(z,Q)$ is the fragmentation function, characterizing the probability of an $i$ parton fragmenting into a hadron, $h$ at a given $z$ and $Q$. The $Q$ evolution of the FF is described by DGLAP evolution equations~\cite{ellis2003qcd},
\begin{align}\label{eq:DGLAP}
\frac{\mathrm{d}D_i^h(z,\mu)}{\mathrm{d}\text{ln}Q^2}=\sum_j\int^1_z\frac{\mathrm{d}x}{x}P_{ij}\left(z/x,Q^2\right)D^h_i\left(x,Q^2\right)
\end{align}
where $i,j$ runs for all (anti)quarks and the gluon. The $P_{ij}$ are the splitting functions from pQCD.

FFs are arbitrary functions satisfying Eq.~\eqref{eq:DGLAP} and $\int^1_0dz\,zD^h_i(z,Q)\!<\!\infty$. The later is required for the momentum conservation and so the probability interpretation. There is no unified theory to determine FFs, thus phenomenological ansatz is used for the form and then it is parametrized with experimental data~\citep{Patrignani:2016xqp,Bertone:2018ecm}. From Eq.~\eqref{eq:differential_fragmentation} one can see that the shape of the hadron spectrum is the result of the convolution integral with the arbitrary shaped FF. Therefore, the spectra is not a pure power law despite the frequently referred statement about pQCD.

The most common FF ansatz is polynomial, with $Q$ dependent parameters, $N^h_i,\alpha^h_i$ and $\beta^h_i$,
\begin{equation}\label{eq:poly_fragmentation}
D^h_i(z,Q)=N^h_iz^{\alpha^h_i}(1-z)^{\beta^h_i}.
\end{equation}
Previous studies have determined the parameters of this FF and it uncertainties with experimental data~\citep{Hirai:2007cx,Kniehl:2000fe, deFlorian:2015}. This form above works perfectly above 5 GeV/c, however its parameters differ from theoretical expectations, based on the parton splittings~\citep{Patrignani:2016xqp}. 

In our previous studies, we determined the parameters of the Tsallis distribution in cases of various spectra for $e^-e^+,pp,pA,AA$ collisions, different hadrons, and kinematic regimes~\cite{Biro:2017arf}. We determined $q\approx1.05-1.25$ for $pp$ at LHC energies, increasing with the collision energy. The temperature-like parameter $T/Q\approx0.05-0.2$, weakly depends on the collision energy and in contrast to $q$, it is independent from the number of valence quarks. We developed a non-extensive jet fragmentation function, the so-called microcanonical Tsallis distribution in Ref.~\citep{Urmossy:2011xk} and we postulate the result as an FF ansatz,
\begin{equation}\label{eq:microTstallis}
D^h_i(z,Q)=N^h_i(1-z)\left[1-\frac{q^h_i-1}{T^h_i}\log(1-z)\right]^{-\frac{1}{q^h_i-1}},
\end{equation}
where $N^h_i,q^h_i,T^h_i$ are $Q$ dependent parameters, thus it has the same number of parameters as Eq.\eqref{eq:poly_fragmentation}.

\section{Results}

To obtain the parameters of our FF, we defined Eq.~\eqref{eq:microTstallis} at a given $Q_0$, then we evolved to $Q$ using Eq.~\eqref{eq:DGLAP} to minimize the $\chi^2_{red}$ function, by changing the parameters of the FF,
\begin{equation}\label{eq:chi}
\chi^2_{red}=\sum_i\frac{(F^h(z_i,Q)-y^h(z_i,Q))^2}{(\delta y^h(z_i))^2N^h_{dof}},
\end{equation}
where $y$ and $\delta y$ is the measured spectrum value and its uncertainty, and $N_{dof}=15$ is the number of parameters. We restricted our parameters to be positive, but no further constrains were added, counter to the studies using Eq.~\eqref{eq:poly_fragmentation}. In this paper, we use $e^-e^+\rightarrow \pi^\pm$ data at $\sqrt s\!=\!91.2$ GeV, from Ref.~\citep{Hirai:2007cx}. We follow all notations and definitions from Ref.~\citep{Hirai:2007cx} in leading order (LO), e.g. independent channels or initial scales $Q_0\!=\!1$ GeV for $u,d,s$ and $g$, $Q_0\!=\!m_c,m_b$ for $c$ and $b$ respectively.

\begin{figure}[H]
\centering
\includegraphics[width=0.51\linewidth]{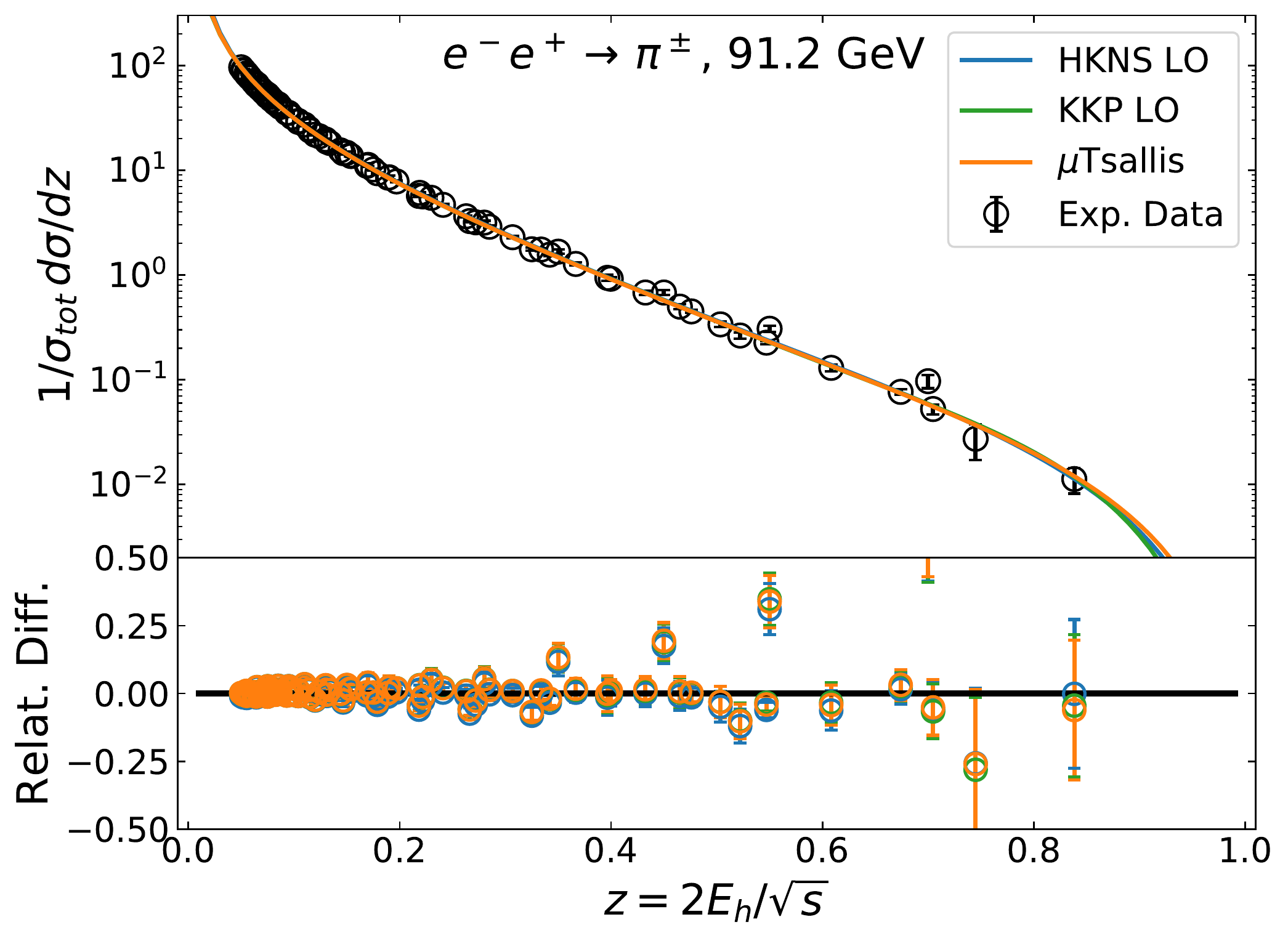}
\includegraphics[width=0.48\linewidth]{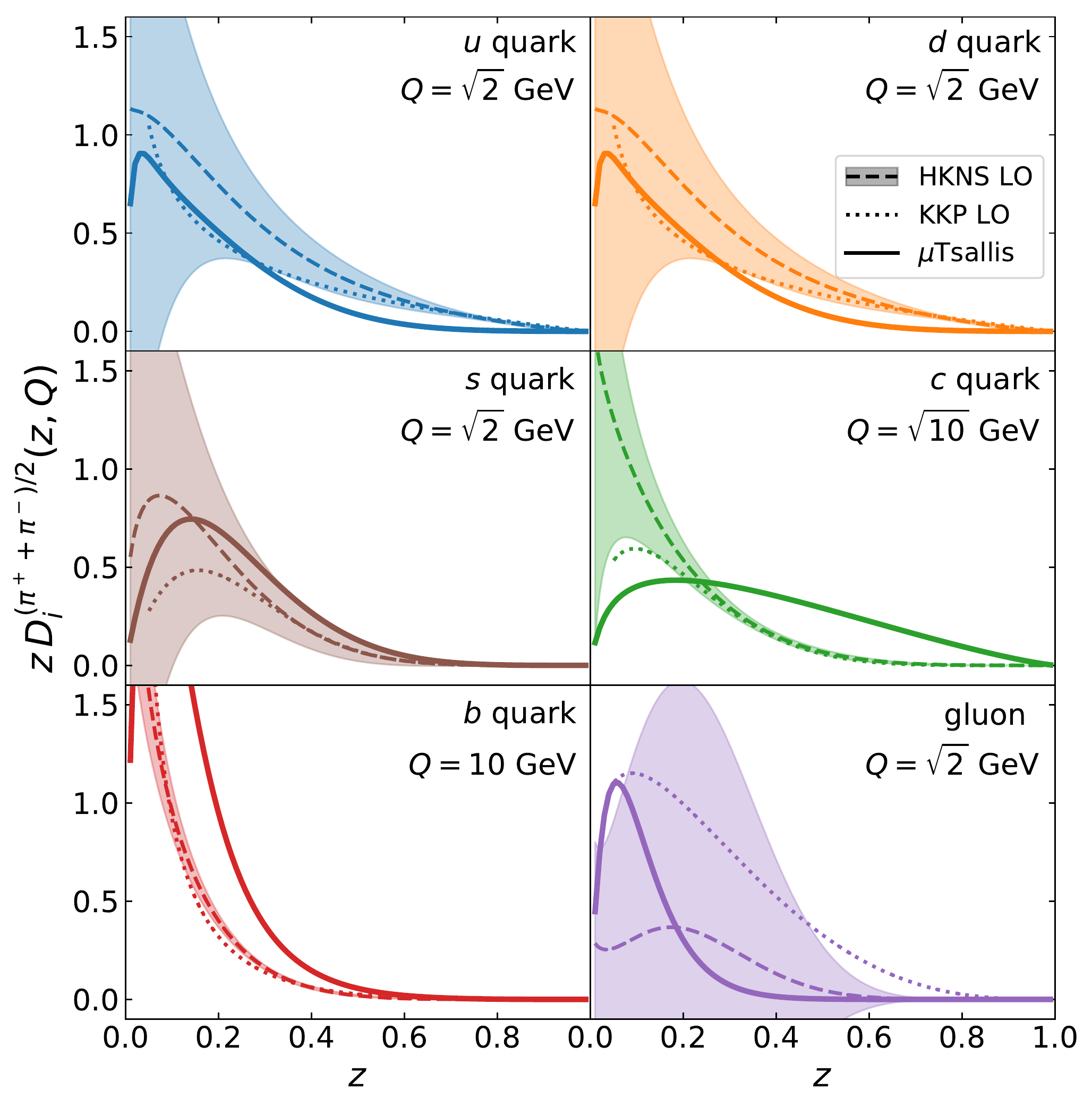}
\caption{\textit{Left}: The microcanonical Tsallis FF parametrization in pion spectrum from  Ref.~\citep{Hirai:2007cx}. HKNS and KKP labels two LO calculations using polynomial FF from Ref.~\citep{Hirai:2007cx,Kniehl:2000fe}. Agreement between different models is clear. \textit{Right}: The microcanonical Tsallis FF parametrization compared to HKNS with uncertainty and KKP FFs for different partonic channels at a given $Q$.
\label{fig:ee_to_pion_total}
}
\end{figure}

We summarize our first results from minimizing Eq~\eqref{eq:chi}. The left panel of Fig.~\ref{fig:ee_to_pion_total} shows the measured spectrum and the fit we got. As a comparison, we plotted calculations with the HKNS~\citep{Hirai:2007cx} and KKP~\citep{Kniehl:2000fe} parametrizations in LO using Eq.~\eqref{eq:poly_fragmentation}. The smallness of the relative differences highlight that our result is in a good agreement with the data and that there is almost no substantial difference between the various models. The right panel of Fig.~\ref{fig:ee_to_pion_total} shows the probability densities, $zD^h_i(z,Q)$ of the fitted microcanonical Tsallis FFs, compared to the HKNS with uncertainties and KKP results at a given $Q$. The different partonic channels agree mostly, however our formula does not diverge at $z=0$ and its slopes are different. For simplicity, we did not use flavor tagged data in the fit, causing the differences seen in the $c$ and $b$ channels.

\begin{figure}[H]
\centering
\includegraphics[width=0.32\linewidth]{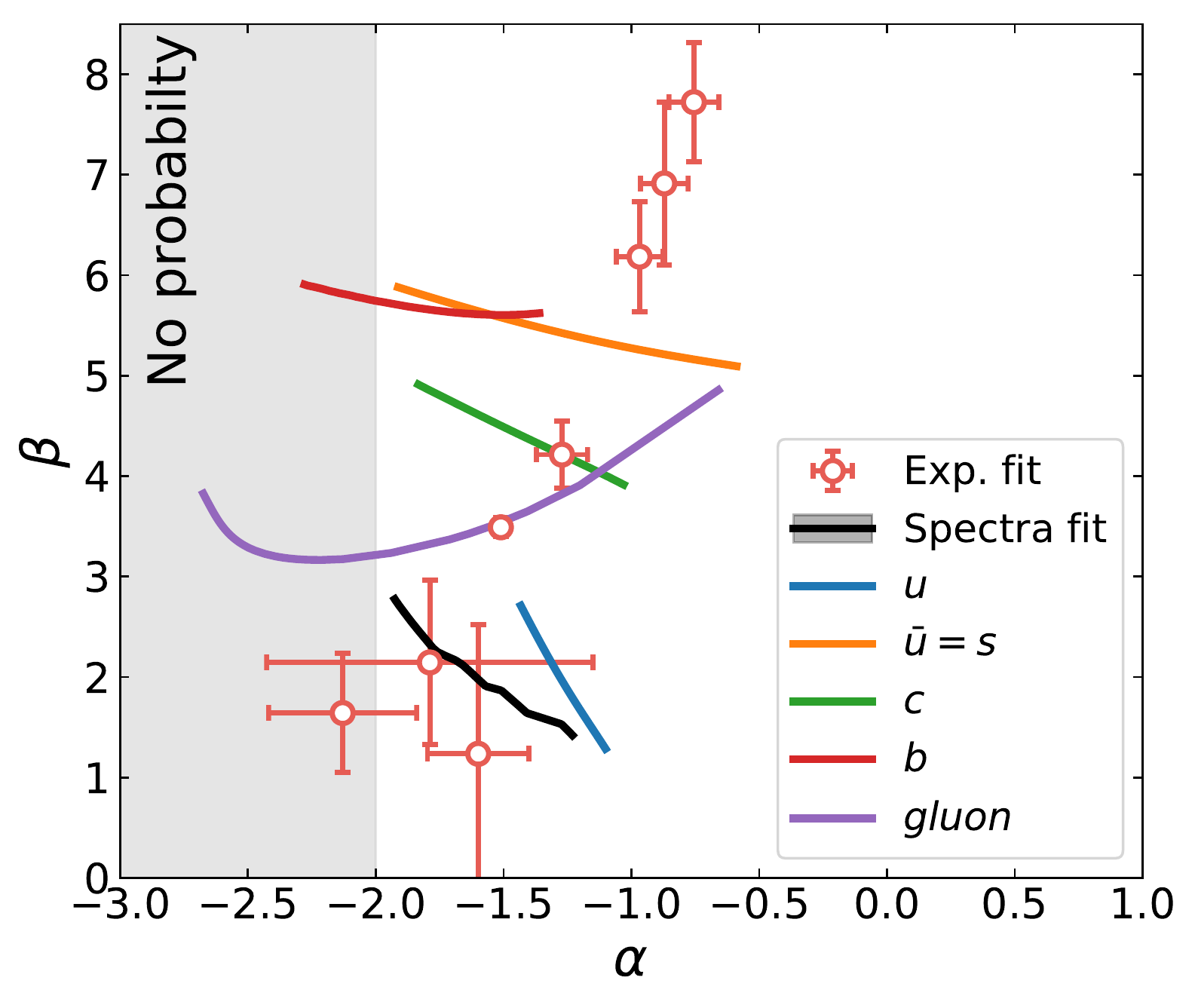}
\includegraphics[width=0.32\linewidth]{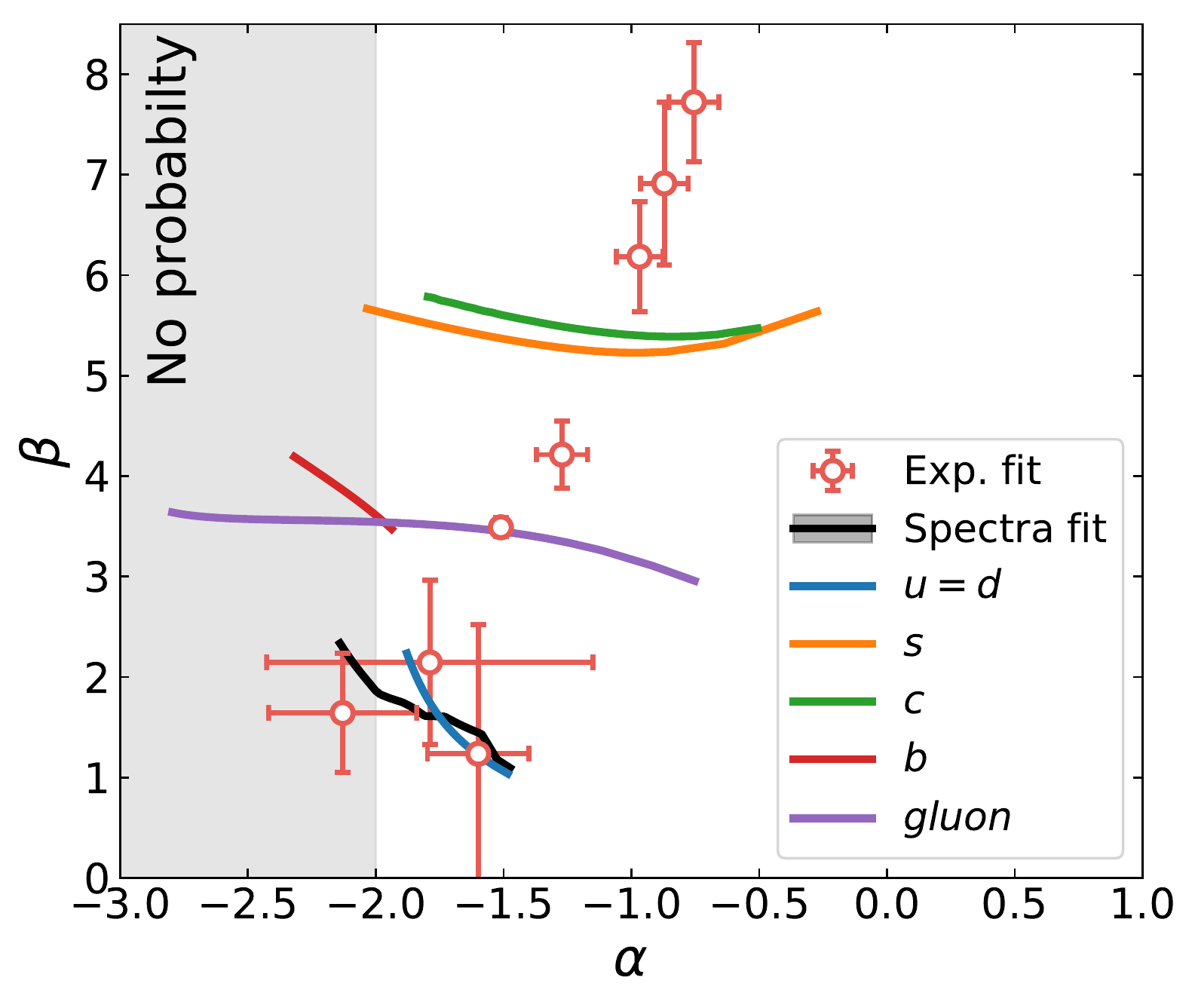}
\includegraphics[width=0.335\linewidth]{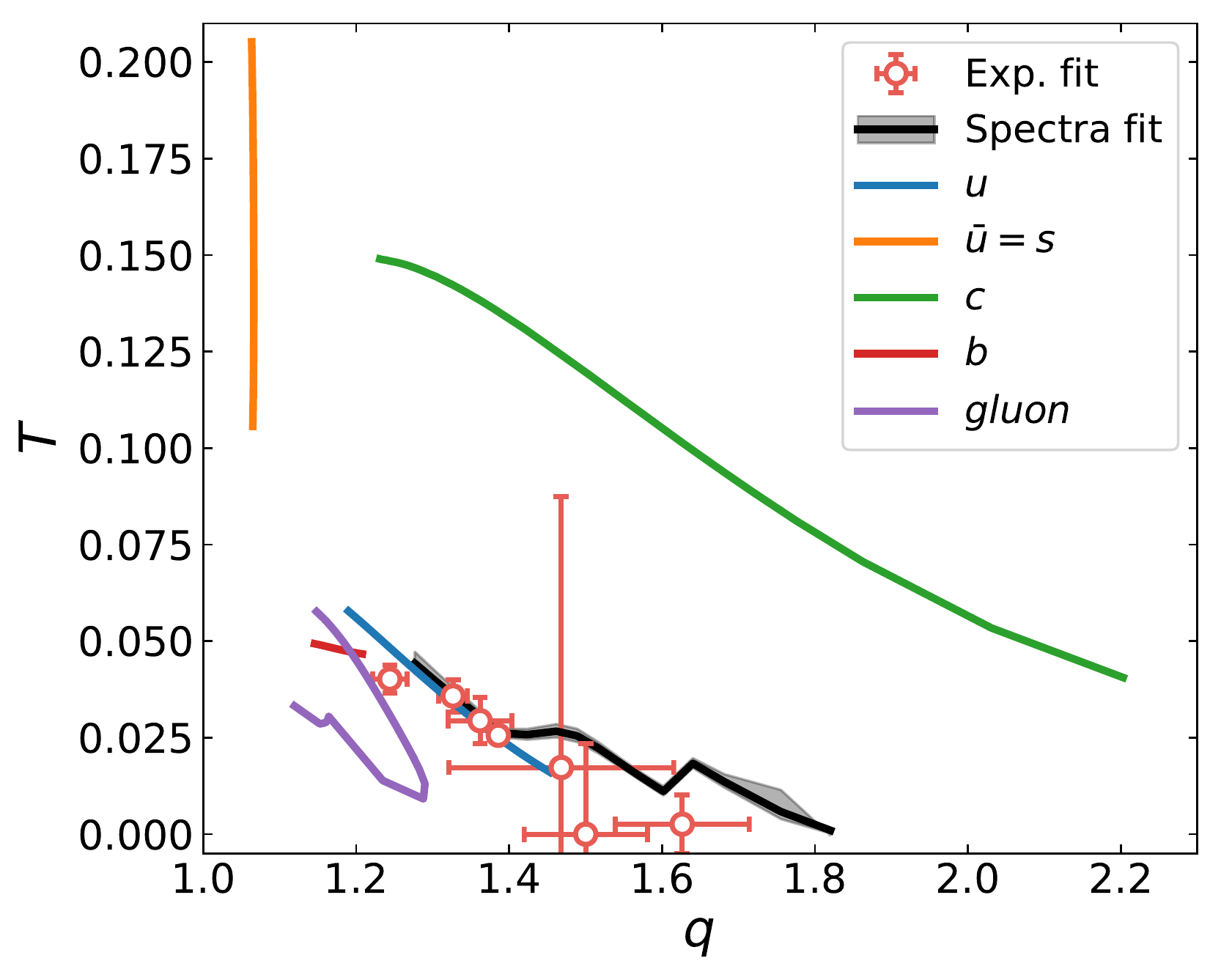}
\caption{Parameters of the HKNS (\textit{left}), KKP (\textit{middle}) and microcanonical Tsallis FF (\textit{right}) are plotted for different $Q$ values and partonic channels. Gray area shows where the polynomial FFs loose the probability meaning. Experimental data and the scale evolved spectra was also fitted with a single distribution in respect to Eq.~\eqref{eq:poly_fragmentation} and \eqref{eq:microTstallis}, described in the text.
\label{fig:ee_to_pion_parameters}
}
\end{figure} 

Fig.~\ref{fig:ee_to_pion_parameters} shows FF parameters $\alpha^{\pi^+}_i, \beta^{\pi^+}_i$ from Eq.~\eqref{eq:poly_fragmentation} for HKNS (left) and KKP (middle) parametrization, and $q^{\pi^+}_i,T^{\pi^+}_i$ parameters from Eq.~\eqref{eq:microTstallis} microcanonical Tsallis FF (right) for different $Q$ between $\sqrt s=2-10^5$ GeV. If $\alpha^h_i\!<\!-2$, then $\int dz\,zD^h_i\rightarrow\infty$ diverges, therefore FF looses the probability interpretation (gray area on Fig.~\ref{fig:ee_to_pion_parameters}). With Eq.~\eqref{eq:DGLAP}, the energy evolution of the pion spectrum from Eq.~\eqref{eq:total_fragmentation} was determined for all three cases. We fitted these spectra with a single function using Eq.~\eqref{eq:poly_fragmentation} for HKNS and KKP and Eq.~\eqref{eq:microTstallis} for microcanonical Tsallis. On the HKNS and KKP panels one can see that there is a weak correlation between the partonic parameters and parameters obtained from the spectrum fits. However, in the case of the microcanonical Tsallis, the correlation is stronger. Moreover, in all cases, the FF of $u$-quark agrees best with the spectra fit, which was expected because of their dominating contribution in pions. For more comparison, we fitted some further experimental data with the same way between $\sqrt s=4.8-91.2$ GeV from Ref.~\cite{Hirai:2007cx}, which overlapped with the microcanonical Tsallis description.

\section{Conclusion}

In this paper we performed a non-extensive motivated fragmentation function parametrization and we calculated the LO spectra of pions in high-energy $e^-e^+$ collision. We compared and verified our FF with widely used other FF parametrizations and experimental data as well. We obtained that our microcanonical Tsallis FF differs from the HKNS~\cite{Hirai:2007cx} and KKP~\cite{Kniehl:2000fe} FFs at lower energy scales and momenta, which could affect the low momentum calculations of proton-proton collisions. The attractor behaviour of the DGLAP evolution could ensure the same behaviour at high $Q$~\cite{Blaizot:2015jea}, supporting the power law tailed pQCD observations. Our fragmentation suggests that there could be a more suitable description for the fragmentation, which overlaps better with the experimental data.


\vspace{6pt} 





\funding{Supported by ÚNKP-18-3 New National Excellence Program of the Ministry of Human Capacities, OTKA K120660, K123815, THOR COST action CA15213, Wigner Data Center and Wigner GPU Laboratory.}

\acknowledgments{We would like to thank for Gergely Kalmar and the HKNS group for sharing their software.}

\reftitle{References}


\externalbibliography{yes}
\bibliography{main}



\end{document}